\documentclass[a4paper, 11pt]{amsart}
\usepackage{amssymb,array}
\usepackage{url}
\usepackage{graphicx}
\usepackage{subfigure}

\usepackage[left=3cm, right=3cm, top=3cm, bottom=3cm]{geometry}

\renewcommand{\leq}{\leqslant}
\renewcommand{\geq}{\geqslant}

\newcommand{\C}{\mathbb{C}}
\newcommand{\E}{\mathbb{E}}

\renewcommand{\H}{\mathcal{H}}
\newcommand{\K}{\mathcal{K}}
\renewcommand{\L}{\mathcal{L}}
\newcommand{\U}{\mathcal{U}}
\newcommand{\D}{\mathcal{D}}
\newcommand{\M}{\mathcal{M}}

\renewcommand{\S}{\mathcal{S}}

\newcommand{\ol}{\overline}

\newcommand{\bra}[1]{\langle #1 |}
\newcommand{\ket}[1]{| #1 \rangle}

\newcommand*{\invstackrel}[2]{\mathop{#1}\limits_{#2}}

\newcommand{\Bell}{\mathrm{Bell}}

\DeclareMathOperator{\trace}{Tr}

\DeclareMathOperator{\I}{I}
\DeclareMathOperator{\id}{id}
\DeclareMathOperator{\Wg}{Wg}
\DeclareMathOperator{\Mob}{Mob}

\DeclareMathOperator{\End}{End}
\DeclareMathOperator{\Rem}{Rem}

\newcommand{\isom}{\simeq}
\newcommand{\scalar}[2]{\langle #1 , #2\rangle}

\renewcommand{\phi}{\varphi}
\newcommand{\iy}{\infty}

\newcommand{\ketbra}[2]{| #1 \rangle \langle #2 |}

\newtheorem{theorem}{Theorem}[section]
\newtheorem{definition}[theorem]{Definition}
\newtheorem{proposition}[theorem]{Proposition}

\newtheorem{lemma}[theorem]{Lemma}

\begin{document}

\title[Random quantum channels I]{Random quantum channels I: graphical calculus
and the Bell state phenomenon}
\author{Beno\^{\i}t Collins}
\address{
D\'epartement de Math\'ematique et Statistique, Universit\'e d'Ottawa,
585 King Edward, Ottawa, ON, K1N6N5 Canada
and 
CNRS, Institut Camille Jordan Universit\'e  Lyon 1, 43 Bd du 11 Novembre 1918, 69622 Villeurbanne, 
France} 
\email{bcollins@uottawa.ca}
\author{Ion Nechita}
\address{Universit\'e de Lyon, Institut Camille Jordan, 43 blvd du 11 novembre 1918, F-69622 Villeurbanne cedex, France} 
\email{nechita@math.univ-lyon1.fr}

\begin{abstract}
This paper is the first of a series where we study 
quantum channels from the random matrix point of view. 
We develop a graphical tool that allows us to 
compute the expected moments of the output of a random quantum channel.

As an application, we study variations of random matrix models
introduced by Hayden \cite{hayden}, and show that their eigenvalues
converge almost surely.

In particular we obtain for some models sharp improvements on the value of 
the largest eigenvalue, and this is shown in a further work to have
new applications to minimal output entropy inequalities.
\end{abstract}

\maketitle

\section{Introduction, motivation \& plan}

The theory of random matrices is a field of its own,
 but whenever it comes to applications,
the driving idea is almost always that although it is very difficult to exhibit matrices having specified
 properties, a suitably chosen random matrix will have very similar properties as the original matrix
 with a high probability. 
 This idea has been used successfully for example in operator algebra with
 the theory of free probability.
 
 In 2007 for the first time, a similar leitmotiv was used with success by Hayden in \cite{hayden}
 and Hayden-Winter in \cite{hayden-winter}
to disprove the R\'enyi entropy additivity conjecture for a wide class of parameters $p$. 
A proof for the most important case $p=1$ was even 
announced by Hastings in \cite{hastings} with probabilistic arguments of different nature. 
 This is arguably the most important conjecture in quantum information theory,
 and the random matrix models introduced by Hayden and their modifications due
 to Hastings seemed very new from our random matrix points of view.
 
This paper is therefore an attempt to understand these matrix models with random matrix techniques. 
For this purpose, we introduce a formalism that is very close to that of planar algebras of Jones \cite{jones},
and we suggest that 
 the language of quantum gates and planar algebras should be considered
 as very closely related to each other. 
 
 Our paper is organized as follows.
In Section \ref{sec:weingarten} we recall known facts about integration over unitary groups
and their large dimension asymptotics. This is nowadays known as Weingarten calculus.
In Section \ref{sec:graphic-model}, we introduce a graphical model to represent (random) matrices arising in random quantum calculus.
Section 
\ref{sec:planar-expansion}
gives a theoretical method for computing expectations with our graphical model
and in the last two sections we give explicit applications of these techniques to random quantum channels. More precisely, in Section \ref{sec:product_independent} we investigate tensor products of two independent quantum channels and in Section \ref{sec:product_conjugate} we look at a product of a random channel $\Phi^U$ with the channel $\Phi^{\ol U}$ defined by the conjugate unitary $\ol U$.

Limit theorems presented in this paper are just a sample of what we can be accomplished with the calculus developed in Section \ref{sec:planar-expansion}. New results will be given in the forthcoming papers \cite{CN2, CN3}.

\section{Background on Weingarten calculus and quantum channels}

\subsection{Weingarten calculus}
\label{sec:weingarten}
This section contains some basic material on unitary integration and Weingarten calculus. A more complete exposition of these matters can be found in \cite{collins-imrn, collins-sniady}. We start with the definition of the Weingarten function.

\begin{definition}
The unitary Weingarten function 
$\Wg(n,\sigma)$
is a function of a dimension parameter $n$ and of a permutation $\sigma$
in the symmetric group $\S_p$. 
It is the inverse of the function $\sigma \mapsto n^{\#  \sigma}$ under the convolution for the symmetric group ($\# \sigma$ denotes the number of cycles of the permutation $\sigma$).
\end{definition}

Notice that the  function $\sigma \mapsto n^{\# \sigma}$ is invertible as $n$ is large, as it
behaves like $n^p\delta_e$
as $n\to\infty$.
If $n<p$ the function is not invertible any more, but we can
keep this definition by taking the pseudo inverse 
and the theorems below will still hold true
(we refer to \cite{collins-sniady} for historical references and further details). We shall use the shorthand notation $\Wg(\sigma) = \Wg(n, \sigma)$ when the dimension parameter $n$ is obvious.

The function $\Wg$  is used to compute integrals with respect to 
the Haar measure on the unitary group (we shall denote by $\U(n)$ the unitary group acting on an $n$-dimensional Hilbert space). The first theorem is as follows:

\begin{theorem}
\label{thm:Wg}
 Let $n$ be a positive integer and
$\mathbf{i}=(i_1,\ldots ,i_p)$, $\mathbf{i'}=(i'_1,\ldots ,i'_p)$,
$\mathbf{j}=(j_1,\ldots ,j_p)$, $\mathbf{j'}=(j'_1,\ldots ,j'_p)$
be $p$-tuples of positive integers from $\{1, 2, \ldots, n\}$. Then
\begin{multline}
\label{bid} \int_{\U(n)}U_{i_1j_1} \cdots U_{i_pj_p}
\overline{U_{i'_1j'_1}} \cdots
\overline{U_{i'_pj'_p}}\ dU=\\
\sum_{\sigma, \tau\in \S_{p}}\delta_{i_1i'_{\sigma (1)}}\ldots
\delta_{i_p i'_{\sigma (p)}}\delta_{j_1j'_{\tau (1)}}\ldots
\delta_{j_p j'_{\tau (p)}} \Wg (n,\tau\sigma^{-1}).
\end{multline}

If $p\neq p'$ then
\begin{equation} \label{eq:Wg_diff} \int_{\U(n)}U_{i_{1}j_{1}} \cdots
U_{i_{p}j_{p}} \overline{U_{i'_{1}j'_{1}}} \cdots
\overline{U_{i'_{p'}j'_{p'}}}\ dU= 0.
\end{equation}
\end{theorem}

Since we shall perform integration over large unitary groups, we are interested in the values of the Weingarten function in the limit $n \to \iy$. The following result encloses all the data we need for our computations
about the asymptotics of the $\Wg$ function; see \cite{collins-imrn} for a proof.

\begin{theorem}\label{thm:mob} For a permutation $\sigma \in \S_p$, let $\text{Cycles}(\sigma)$ denote the set of cycles of $\sigma$. Then
\begin{equation}
\Wg (n,\sigma )=(-1)^{n-\# \sigma}
\prod_{c\in \text{Cycles} (\sigma )}\Wg (n,c)(1+O(n^{-2}))
\end{equation}
and 
\begin{equation}
\Wg (n,(1,\ldots ,d) ) = (-1)^{d-1}c_{d-1}\prod_{-d+1\leq j \leq d-1}(n-j)^{-1}
\end{equation}
where $c_i=\frac{(2i)!}{(i+1)! \, i!}$ is the $i$-th Catalan number.
\end{theorem}

A shorthand for this theorem is the introduction of a function $\Mob$ on the symmetric
group, invariant under conjugation and multiplicative over the cycles, satisfying
for any permutation $\sigma \in \S_p$:
\begin{equation}
\Wg(n,\sigma) = n^{-(p + |\sigma|)} (\Mob(\sigma) + O(n^{-2})).
\end{equation}
where $|\sigma |=p-\# \sigma $ is the \emph{length} of $\sigma$, i.e. the minimal number of transpositions that multiply to $\sigma$. We refer to \cite{collins-sniady} for details about the function $\Mob$.
We finish this section by a well known lemma which we will use several times towards the end of the paper. This result is contained in \cite{nica-speicher}.
\begin{lemma}\label{lem:S_p}
The function
$d(\sigma,\tau) = |\sigma^{-1} \tau|$ is an integer valued distance on $\S_p$. Besides, it has the following properties:
\begin{itemize}
\item the diameter of $\S_p$ is $p-1$;
\item $d(\cdot, \cdot)$ is left and right translation invariant;
\item for three permutations $\sigma_1,\sigma_2, \tau \in \S_p$, the quantity $d(\tau,\sigma_1)+d(\tau,\sigma_2)$
has the same parity as $d(\sigma_1,\sigma_2)$;
\item the set of geodesic points between the identity permutation $\id$ and some permutation $\sigma \in \S_p$ is in bijection with the set of non-crossing partitions smaller than $\pi$, where the partition $\pi$ encodes the cycle structure of $\sigma$. Moreover, the preceding bijection preserves the lattice structure. 
\end{itemize}
\end{lemma}

We end this section by the following definition which generalizes the trace function. For some matrices $A_1, A_2, \ldots, A_p \in \M_n(\C)$ and some permutation $\sigma \in \S_p$, we define
\[\trace_\sigma(A_1, \ldots, A_p) = \prod_{\substack{c\in \text{Cycles} (\sigma) \\ c=(i_1 \, i_2 \, \cdots \, i_k)}} \trace \left( A_{i_1} A_{i_2} \cdots A_{i_k}\right).\]
We also put $\trace_\sigma(A) = \trace_\sigma(A, A, \ldots, A)$.
\subsection{Quantum channels}

In Quantum Information Theory, a \emph{quantum channel} is the most general transformation of a quantum system. Quantum channels generalize the unitary evolution of isolated quantum systems to \emph{open quantum systems}. Mathematically, we recall that a quantum channel is a linear completely positive trace preserving map $\Phi$ from $\M_n(\C)$ to itself. The trace preservation condition is necessary since quantum channels should map density matrices to density matrices. The complete positivity condition can be stated as
\[ \forall d \geq 1, \quad \Phi \otimes \I_d : \M_{nd}(\C) \to \M_{nd}(\C) \text{ is a positive map.}\]

The following two characterizations of quantum channels turn out to be very useful.

\begin{proposition}\label{prop:stinespring_kraus}
A linear map $\Phi : \M_n(\C) \to \M_n(\C)$ is a quantum channel if and only if one of the following two equivalent conditions holds.
\begin{enumerate}
\item \textbf{(Stinespring dilation)} There exists a finite dimensional Hilbert space $\K = \C^{d}$, a density matrix $Y \in \M_{d}(\C)$ and an unitary operation $U \in \U(nd)$ such that
\begin{equation}\label{eq:Stinespring_form}
\Phi(X) = \Phi^{U,Y}(X) = \trace_\K\left[ U(X \otimes Y) U^* \right], \quad \forall X \in \M_n(\C).
\end{equation}
\item \textbf{(Kraus decomposition)} There exists an integer $k$ and matrices $L_1, \ldots, L_k \in \M_n(\C)$ such that
\[\Phi(X) = \sum_{i=1}^{k} L_i X L_i^* ,\quad \forall X \in \M_n(\C).\]

and
\[\sum_{i=1}^{k} L_i^* L_i = \I_n.\]
\end{enumerate}
\end{proposition}

It can be shown that the dimension of the ancilla space in the Stinespring dilation theorem can be chosen $d = \dim \K = n^2$ and that the state $Y$ can always be considered to be a rank one projector. A similar result holds for the number of Kraus operators: one can always find a decomposition with $k=n^2$ operators. 

In the final two sections of the paper we study a model of \emph{random quantum channels} originating from the Stinespring dilation formula (\ref{eq:Stinespring_form}). We shall be interested in the spectral properties of the elements in the image of such random channels. Quantum channels will be the main field of application of the graphical calculus we develop in the next two sections, our aim being the treatment of some additivity problems in quantum information theory.

\section{Graphical model}
\label{sec:graphic-model}

In this section, we lay out the foundation for the graphical calculus we shall develop later. We introduce a graphical formalism for representing tensors and tensor contractions that is adapted to quantum information theory. We start at an abstract level, with a purely diagrammatic axiomatization and then we study the Hilbert representations, where graph-theoretic objects shall be associated with concrete elements of Hilbert spaces.

\subsection{Diagrams and tensors} 

\subsection*{Diagrams, boxes, decorations and wires} 
Our starting point is a set $\tilde{S}$ endowed with an involution without fixed point $*$. The set $\tilde{S}$ splits as $S\sqcup S^*$ according to the involution. Elements of $\tilde{S}$ are called \emph{decorations}.

A \emph{diagram} is a collection of decorated \emph{boxes} and possibly \emph{wires} (or strings) connecting the boxes along their decorations according to rules which we shall specify. In terms of graph theory, a diagram is an unoriented (multi-)graph whose vertices are boxes, and whose edges are strings. Each vertex comes with a (possibly empty) $n$-tuple of indices (or decorations or labels) in $\tilde{S}^n$. The number $n$ of decorations may depend on the vertex. We say that two diagrams are isomorphic if they are isomorphic as multi-graphs with labeled vertices.

A box is an elementary diagram from which we can construct more elaborate diagrams by putting boxes together and possibly wiring them together. Each box $B$ of a diagram has attached to it a collection of $n(B)$ decorations in $\tilde{S}^{n(B)}$. The union of the decorations attached to a box $B$ is denoted by $S(B)\sqcup S^*(B)$.

\begin{figure}[ht]
\includegraphics{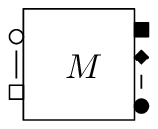}
\caption{A box $M$}
\label{fig:box}
\end{figure}

Graphically, boxes are represented by rectangles with symbols corresponding to the decorations attached to them (see Figure \ref{fig:box}). We take the convention that decorations in $S^*$ are represented by empty (or white) symbols and decorations in $S$ by full (or black) symbols. Each decoration is thought as having potentially up to two \emph{attachment points}. An \emph{inner} one (which is attached to the box it belongs to) and an \emph{outer} one, which we shall allow to be attached to a string later on.

\subsection*{Constructing new diagrams out of old ones}
Given a family of existing diagrams (e.g. boxes) there exists several ways of creating new diagrams.
\begin{enumerate}
\item One can put diagrams together, i.e. take their disjoint union
 (when it comes to taking representations 
in Hilbert spaces, this operation will amount to tensoring). One diagram can be viewed as a box. This amounts to specifying an order between the boxes.
\item Given a diagram $A$ and a complex number $x$, one can create a new diagram $A'=xA$.

\item Given two boxes $A,B$ having the same $n$-tuples of decorations, one can define $A+B$. This axiom and the previous one  (together with evident relations such as $A+A=2A$ which we don't enumerate in detail) endow the set of identically decorated diagrams with a structure of a complex vector space. 

\item One can add \emph{wires} to an existing diagram (or between
two diagrams that have been put together). A wire is allowed between 
the outer attachment of two decorations only if the decorations have the same shape and different shadings. Such a wire can be created if and only if the two candidate decorations have their outer attachments unoccupied.

\item There exists an anti-linear 
 \emph{involution}
 on the diagrams, denoted by $*$.
 This operation does nothing on the wires. On the boxes, it reverts the shading of the decorations. The involution $*$ is conjugate linear.
\end{enumerate}

\subsection*{Hilbert structure}

We shall now consider a concrete representation of the diagrams introduced above as tensors in Hilbert spaces. We start by assuming that the set $S$ of full (or black) decorations corresponds to a collection of finite dimensional Hilbert spaces $S=\{V_1, V_2, \cdots \}$. An important fact that will be useful later is that each Hilbert space $V_i$ comes equipped with an orthonormal basis $\{e_1, e_2, \ldots, e_{\dim V_i}\}$. Our aim is to define a $*$-linear map $T$ between the diagrams and tensors in products of Hilbert spaces in the above class and their duals. By duality, white decorations correspond to dual spaces $S^*=\{V_1^*, V_2^*, \cdots \}$. With these conventions, boxes can be seen as tensors whose legs belong to the vector spaces corresponding to its decorations. In a diagram, symbols of the same shape denote isomorphic spaces, but the converse may be false. A particular space $V_i$ (or $V_i^*$) can appear several times in a box. The reader acquainted with quantum mechanics might think of white shapes as corresponding to `bras' and black shapes corresponding to `kets', but we shall get back to quantum mechanical notions later. 

To a box $B$ we therefore associate a tensor
\begin{equation}
\label{eq:def-TonBox}
T_B \in \left[\bigotimes_{i \in S(B)} V_i\right] \otimes \left[ \bigotimes_{j \in S^*(B)} V_j^* \right].
\end{equation}
Using the canonical duality between tensors and linear applications, $T_B$ can also be seen as a linear map
\[ T_B : \bigotimes_{j \in  S^*(B)} V_j \to \bigotimes_{i \in S(B)} V_i,\]
We use freely partial duality results, and for example, an element of $V \otimes W^*$ can as well be seen as an element of $\mathcal L(W,V)$ or $\mathcal L(V^*,W^*)$. 

Equation (\ref{eq:def-TonBox}) defines the map $T$ from the collection of boxes to the collection of vectors in Hilbert spaces obtained by tensoring finitely many copies of $V_i,i\in S(B)\cup S^*(B)$. This map is denoted by
\[T:B\mapsto T_B\]
and we now explain how we can extend it to all diagrams. A \emph{wire} connecting two decorations of the same shape (corresponding to some Hilbert space $V$) is associated with the identity map (or tensor) $\I : V \to V$. Together with our duality axiom, it also corresponds to a canonical tensor contraction (or trace)
\[ C : V^* \otimes V \to \C.\]
We denote the set of wires in a diagram $\D$ by $\mathcal C(\D)$.

With this notation, a diagram $\D$ is associated with the tensor $T$ obtained by applying all the contractions (``wires'') to the product of tensors represented by the boxes. One is left with a tensor 
\[T_\D = \left[ \prod_{C \in \mathcal C(\D)} C \right] \left(\bigotimes_{B \text{ box of } \D} T_B\right).\]
This is well defined (provided that one specifies one total order on the boxes): 
the order of the factors in the product does not matter, since wires act on different spaces. For a box $B$, we denote by ${FS}(B) \subset S(B)$ the subset of black decorations which have no wires attached (we call such a decoration \emph{free}). ${FS}^*(B)$ is defined in the same manner for white decorations (dual spaces). With this notation, the tensor $T_\D$ associated to a diagram $\D$ can be seen in two ways: as an element of a Hilbert space
\[ T_\D \in \left[\bigotimes_{j \in \bigcup_B {FS}^*(B)} V_j^*\right] \otimes \left[ \bigotimes_{i \in \bigcup_B {FS}(B)} V_i \right],\]
or, equivalently, as a linear map
\[ T_\D : \bigotimes_{j \in \bigcup_B {FS}^*(B)} V_j \to \bigotimes_{i \in \bigcup_B {FS}(B)} V_i .\]

We need two further axioms to ensure that we are indeed dealing with acceptable Hilbert representations.
\begin{enumerate}
\item A diagram such that all outer attachments of its decorations are occupied by wires corresponds canonically to an element in $\C$. In addition, a trivial box with a given decoration of type $i$ closed on itself by a wire into a loop takes a value in $\mathbb{N}$. This value is called the dimension of $V_i$.
\item Given a diagram $\D$, if it is canonically paired to its dual $\D^*$ by strings, the result lies in $\mathbb{R}^+$.
\end{enumerate}

\subsection*{Special diagrams.}
To make our calculus useful, we need to introduce a few special diagrams 
(equivalently, boxes) satisfying some specific axioms.
\begin{enumerate}
\item {\bf The trivial box.} A wire connecting two identically shaped decorations of different shading corresponds to the identity map $\I : V \to V$. We shall call this box the \emph{trivial} or the identity box.

\begin{figure}[ht]
\includegraphics{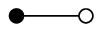}
\caption{Trivial box}
\label{fig:trivial-box}
\end{figure}

It satisfies the following identity axiom:

\begin{figure}[ht]
\includegraphics{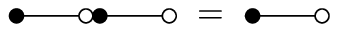}
\caption{Trivial axiom}
\label{fig:trivial-axiom}
\end{figure}

\item {\bf Bras and kets.} The simplest boxes one can consider are vectors and linear forms. Following the quantum mechanics `bra' and `ket' vocabulary,
vectors, or (1, 0)-tensors have no white decorations and only one black decoration, whereas linear forms (or (0,1)-tensors) have one white label and no black labels. Since our Hilbert spaces come equipped with fixed basis, we introduce some special notation for the ket $\ket{e_1} = e_1$ and the bra $\bra{e_1} = \scalar{e_1}{\cdot}$ corresponding to the first vector, as in Figure \ref{fig:ket-bra} (the choice of the first element of the basis being of course arbitrary).

\begin{figure}[ht]
\includegraphics{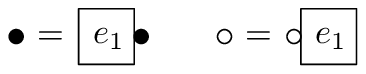}
\caption{Fixed ket and bra}
\label{fig:ket-bra}
\end{figure}

\item {\bf The Bell state.} Since each space $V \in S$ comes equipped  with a particular fixed basis $\{e_i\}_{i=1}^{\dim V}$, we can define the \emph{bra Bell state} as the tensor (it is in fact a linear form)
\[\Bell_V^* = \sum_{i=1}^{\dim V} e_i^* \otimes e_i^*,\]
and its ket counterpart (which is a vector in $V \otimes V$)
\[\Bell_V = \sum_{i=1}^{\dim V} e_i \otimes e_i.\]
This notation is needed in the sense that Bell states are not canonical and are not well defined from the sole data of $V$. They rely on some additional real structure of the vector space $V$ which can be encoded by the data of an explicit basis. Bell states are represented in Figure \ref{fig:Bell:states}. They satisfy the graphical axiom in Figure \ref{fig:Bell:axiom}. Bell states play a central role in our formalism; we shall see later that they allow us to define the \emph{transposition} of a box and even to consider wires connecting identical decorations.

\begin{figure}[ht]
\subfigure[]{\label{fig:Bell:states}\includegraphics{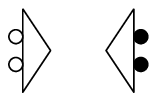}}\quad\quad\quad
\subfigure[]{\label{fig:Bell:axiom}\includegraphics{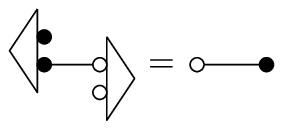}}
\caption{Bell states and axiom}
\label{fig:Bell}
\end{figure}

\item {\bf Unitary boxes.} Boxes associated to unitary matrices $U$ satisfy the graphical axiom depicted in Figure \ref{fig:unitary} which corresponds to the identities $UU^* = U^* U = \I$.

 \begin{figure}[ht]
 \includegraphics{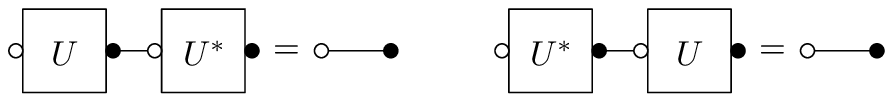}
\caption{Unitary axioms}
\label{fig:unitary}
\end{figure}

\end{enumerate}

\subsection{Examples}

Let us now look at some simple diagrams which illustrate this formalism. 

\begin{figure}[ht]
\subfigure[]{\includegraphics{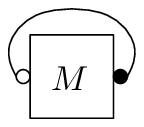}}\quad\quad\quad
\subfigure[]{\includegraphics{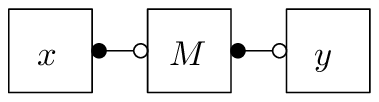}}\quad\quad\quad
\subfigure[]{\includegraphics{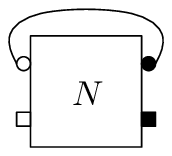}}
\caption{Some simple diagrams}
\label{fig:simple}
\end{figure}

Suppose that each diagram in Figure \ref{fig:simple} comes equipped with two vector spaces $V_1$ and $V_2$ which we shall represent respectively by circle and square shaped symbols. In the first diagram, $M$ is a tensor (or a matrix, depending on which point of view we adopt) $M \in  V_1^* \otimes V_1$, and the wire applies the contraction $V_1^* \otimes V_1 \to \C$ to $M$. The result of the diagram $\D_a$ is thus $T_{\D_a} = \trace(M) \in \C$. In the second diagram, again there are no free decorations, hence the result is the complex number $T_{\D_b} = \scalar{y}{Mx}$. Finally, in the third example, $N$ is a $(2,2)$ tensor or a linear application $N \in \L(V_1 \otimes V_2, V_1 \otimes V_2)$. When one applies to the tensor $N$ the contraction of the couple $(V_1, V_1^*)$, the result is the partial trace of $N$ over the space $V_1$: $T_{\D_c} = \trace_{V_1}(N) \in \L(V_2, V_2)$.

\begin{figure}[ht]
\subfigure[]{\label{fig:Bell_transposition:transposition}\includegraphics{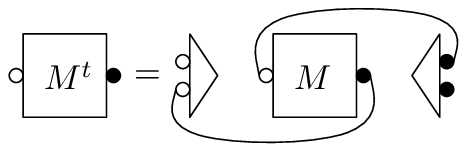}}\quad\quad\quad
\subfigure[]{\label{fig:Bell_transposition:same_color}\includegraphics{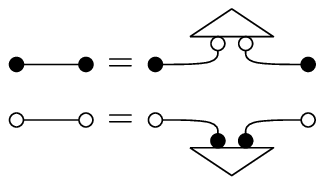}}
\caption{Bell states and transposition}
\label{fig:Bell_transposition}
\end{figure}

Bell states allow us to introduce the \emph{transposition} operation for a tensor (or a box) as follows. We define transposition for a matrix $M$ (or a tensor $M \in V^* \otimes V$) and 
we extend it in a trivial way to more general situations. 
Graphically, the box corresponding to the transposed tensor $M^t$ is defined in Figure \ref{fig:Bell_transposition:transposition}; it consists in connecting an appropriate Bell state to each decoration of the box. Note however that this operation is different from the involution $*$ applied to the same box. Moreover, Bell states allow for wires connecting identical shaped symbols of the same color, as in Figure \ref{fig:Bell_transposition:same_color}. Such non-canonical tensor contractions ($V \otimes V \to \C$ or $V^* \otimes V^* \to \C$) are shorthand graphical notations for  the corresponding diagram containing a Bell state, and we shall use them quite often in what follows. 

\begin{figure}[ht]
\includegraphics{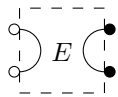}
\caption{The maximal entangled state $E$}
\label{fig:special}
\end{figure}

 Also, for reasons which shall be clear later, we shall sometimes make substitution $\ol M = (M^*)^t$. Finally, by grouping two Bell states together, one obtains the (non-canonical) tensor $E$ (Figure \ref{fig:special}), called ``the maximal entangled state''. It corresponds to the tensor
\[E = \sum_{i=1}^{\dim V} \sum_{j=1}^{\dim V} e_i^* \otimes e_i^* \otimes e_j \otimes e_j \in V^* \otimes V^* \otimes V \otimes V. \]
The reader with background in quantum information will notice that the maximally entangled state we just defined is \emph{not normalized} in order to be a density matrix. The reader with background in planar algebra theory will recognize a multiple of the Jones projection.

\begin{figure}[ht]
\includegraphics{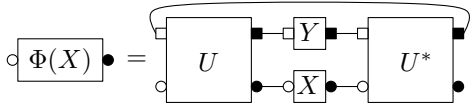}
\caption{Diagram for a quantum channel in its Stinespring form}
\label{fig:quantum_channel}
\end{figure}

The diagram in Figure \ref{fig:quantum_channel} is of crucial importance in what follows. It corresponds to a quantum channel in its Stinespring representation (see Eq. (\ref{eq:Stinespring_form})). Round shaped inputs and outputs correspond to the Hilbert space $\H$ and squares correspond to the ``environment'' $\K$. We shall study such diagrams with random interaction unitaries $U$ in Sections \ref{sec:product_independent} and \ref{sec:product_conjugate}.

\subsection{Comments on other existing graphical calculi}

The above formalism is the one that seemed the most compatible with 
Weingarten calculus. Here, we comment about already existing graphical formalisms,
in the hope that this section will serve as a dictionary for the reader acquainted to one of the calculi below.

Our calculus is mainly inspired by Bob Coecke's \emph{Kindergarten Quantum Mechanics} \cite{coecke}. However we choose not to orient the strings; rather, we separate with color (black/white) the vector spaces and their duals,
therefore there is only one possible pairing. A common feature of the two calculi is the central place occupied in the formalisms by Bell states.

V.F.R. Jones's theory of planar algebras \cite{jones} is also connected to our graphical calculus. One of our diagrammatic axioms is the existence of a Bell state. 
This is very closely related to the axioms of Temperley Lieb algebras and the 
diagrammatic for a Jones projection.
Most of our calculus could take place in Jones' \emph{bipartite graph planar algebra}.

\section{Planar expansion}\label{sec:planar-expansion}

In this section, we consider diagrams that may contain random matrices. This is where probability theory comes into play; we focus on the case where the random elements appearing in the diagrams are random unitary Haar-distributed on some finite dimensional unitary group.

Our task is to compute expectation values $\E(\D )$ of diagrams $\D$ containing boxes associated with random unitary operations. We shall write $\E(\D )$ as a weighted sum of some diagrams obtained from $\D$ that to not contain anymore random tensors.

\subsection{Expectation of a diagram containing random independent unitary matrices}

Suppose we have a diagram $\D$ that has boxes of two types: either boxes of type $U$, $U^*$, $\overline{U}$ or $U^t$ where $U$ is a unitary random variables in a fixed space of type $\End(\otimes_{i\in I}V_i)$, distributed according to the Haar measure on the unitary group $\U(n)$ on this space, or other boxes which are independent (as classical random variables) from $U$ (this includes deterministic boxes or tensors). We shall now present an algorithm for computing the expectation of such a diagram with respect to the probability law of $U$. Before describing the algorithm, let us note that if a diagram contains several \emph{independent} Haar unitary matrices, one can recursively apply the algorithm to compute the expectation over all the random unitary matrices appearing in the diagram.

The first step in our algorithm is to ensure that $\D$ contains only boxes of type $U$ and $\ol{U}$. This can be done by using the transposition transformation via Bell states, and replacing $U^t$ boxes by $U$ boxes with the opposite shading of the decorations and $U^*$ boxes by $\ol U$ boxes.

Next, we introduce a concept of \emph{removal} of boxes $U$ and $\ol U$. A removal $r$ is a procedure which transforms a diagram $\D$ into a new diagram $\D_r$ which does not contain neither $U$ nor $\ol U$ boxes. In other words, $r$ is a way to remove random unitaries $U$ from a diagram $\D$. The set of all admissible removal procedures for a diagram $\D$ will be denoted by $\Rem(\D)$.

We now move on to describe removal procedures and how they operate on diagrams. First of all, a removal is not possible if the number of boxes $U$ in $\D$ is different from the number of boxes $\ol U$. In such a case, the set $\Rem(\D)$ will be defined to be empty. This rule is the diagrammatic equivalent of Eq. (\ref{eq:Wg_diff}) from Theorem \ref{thm:Wg}.

Assuming that the number of $U$ boxes and $\ol U$ boxes is the same, a removal $r$ is a way to pair decorations of the $U$ and $\ol U$ boxes appearing in a diagram. More precisely, $r$ is the data of a pairing $\alpha $ of the white decorations of $U$  boxes with the white decorations of $\ol U$ boxes, together with a pairing $\beta $ between the black decorations of $U$ boxes and the black decorations of $\ol U$ boxes. Assuming that $\D$ contains $p$ boxes of type $U$ and that the boxes $U$ (resp. $\ol U$) are labeled from $1$ to $p$, then $r=(\alpha,\beta)$ where $\alpha,\beta$ are permutations of $\mathcal{S}_p$.

\begin{figure}[ht]
\includegraphics{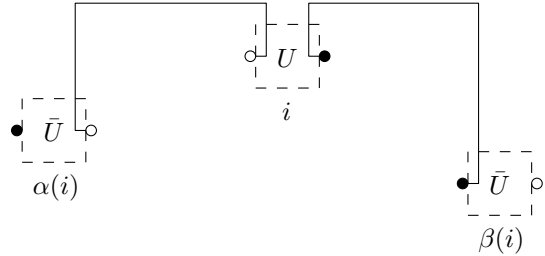}
\caption{Removal: elimination of boxes and pairing of decorations}
\label{fig:expectation_unitary}
\end{figure}

Given a removal $r \in \Rem(\D)$, we construct a new diagram $\D_r$ associated to $r$, which has the important property that it no longer contains boxes of type $U$ or $\ol U$. We proceed in the following way: one starts by erasing the boxes $U$ and $\ol U$ but keeps the decorations attached to them. Assuming that one has labeled the erased boxes $U$ and $\ol U$ with integers from $\{1, \ldots, p\}$, one connects \emph{all} the (inner parts of the) \emph{white} decorations of the $i$-th erased $U$ box with the corresponding (inner parts of the) \emph{white} decorations of the $\alpha(i)$-th erased $\ol U$ box. In a similar manner, one uses the permutation $\beta$ to connect black decorations. A diagrammatic explanation of the above algorithm is described in Figure \ref{fig:expectation_unitary}.

We are now ready to state the main result of this section.

\begin{theorem}\label{thm:Wg_diag}
The following holds true:
\[\E_U(\D)=\sum_{r=(\alpha, \beta) \in \Rem(\D)} \D_r \Wg (n, \alpha\beta^{-1}).\]
\end{theorem}

\begin{proof}
This is just Weingarten calculus of Theorem \ref{thm:Wg} applied to our graphical conventions.
\end{proof}

When more than one independent unitary matrices $U, V, \ldots$ are present in a diagram, we proceed by induction over 
each independent matrix: we successively remove $U$, $V$, etc. One can  check directly that the order of the induction does not change the final result. This is compatible with the probabilistic property of the expectation, $\E_{U,V}(\D)= \E_V(\E_U(\D))$.

Theorem \ref{thm:Wg_diag} might just look as a reformulation of known results, but without this graphical method, obtaining the main results of this paper is extremely cumbersome and very counterintuitive.

Let us now comment on the first step of our removal algorithm, replacing $U^*$ boxes with $\ol U$ boxes. The purpose of such a substitution is purely practical: later in the removal procedure, we pair decorations of \emph{the same color}. If we should have decided to work with $U$ and $U^*$ boxes, one should always pair decorations of different colors, and this can turn out to be rather cumbersome when doing combinatorics. On the other hand, each time we replace a $U^*$ box by a $\ol U$ box, we introduce two more Bell states into our diagram (see Figure \ref{fig:Bell_transposition}); although we decided not to display such states and rather to allow wires connecting decorations of the same color, this operation increases in some sense the ``complexity'' of the diagram.

In the next sub-section we present a warm-up toy example of Theorem
\ref{thm:Wg_diag}.
Further applications of the above theorem will be considered in a forthcoming paper \cite{CN3}, where problems from free probability theory will be treated using similar techniques.

\subsection{First example: partial tracing a randomly rotated matrix}

\begin{figure}
\includegraphics{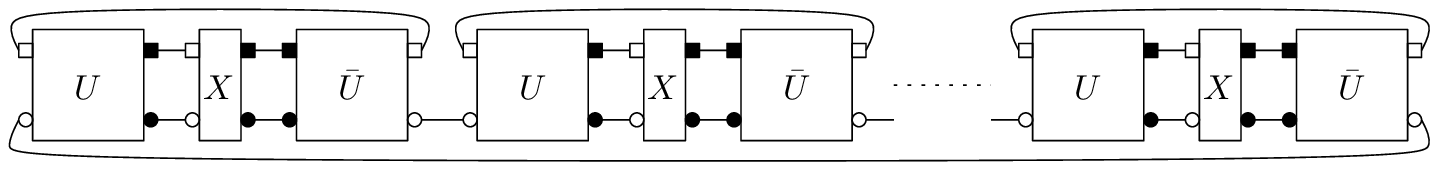}
\caption{The diagram for $\trace(Y^p)$}
\label{fig:random_partial_trace}
\end{figure}

As a first application of the graphical formalism, we consider the following problem. Let $X \in \M_{nk}(\C)$ be a deterministic matrix. In a manner similar to random quantum channels, we define, for a fixed integer parameter $k \geq 1$, the random matrix
\[Y = \trace_k[U X U^*] \in \M_n(\C),\]
where $U \in \U(nk)$ is a Haar distributed random unitary matrix. Notice that we are considering here non-normalized traces. Using our graphical formalism, we shall compute the moments of $Y$, $\E[\trace(Y^p)]$ for all $p \geq 1$. After replacing $U^*$ boxes with $\ol U$, one gets the diagram in Figure \ref{fig:random_partial_trace}, where round decorations correspond to $\C^n$ and square ones to $\C^k$. Note that in Figure \ref{fig:random_partial_trace} there are $p$ groups ``$UX\ol U$'' wired together.

By Theorem \ref{thm:Wg_diag}, the expectation (with respect to the Haar measure of the unitary group $\U(nk)$) of this diagram is a weighted sum (with Weingarten weights) of diagrams $\D_r$ obtained after the removal of $U$ and $\ol U$ boxes. Such diagrams $\D_r$ contain only $X$ blocks and loops of different types. Let us compute the the value of a diagram $\D_r$, where the removal is given by $r=(\alpha, \beta) \in \S_p^2$. 

Suppose we number the boxes from $1$ to $p$ and the permutations $\alpha$ connects the white decorations and $\beta$ the black decorations. The point here is that the permutation $\alpha$ will be responsible for the loops appearing in $\D_r$, whereas $\beta$ will connect $X$ boxes. We start by counting the number of loops in $\D_r$. Loops are of two types, the ones containing the top, square decorations (a loop of this type has a value of $k$) and the ones that come from the bottom, round decorations, each with a value of $n$. Since the top decorations are already connected (in the original diagram $\D$) by the identity permutation, the number of loops of the first type is given by the number of cycles of $\alpha$, $\# \alpha$; they give a total contribution of $k^{\# \alpha}$. Loops corresponding to round decorations are initially connected by the cycle 
\begin{equation}\label{eq:gamma_cycle}
\gamma = ( n \; n-1 \; \cdots 3 \; 2 \; 1) \in \S_p.
\end{equation}
Hence, the number of such loops is $\#(\gamma^{-1} \alpha)$ and they give a total contribution of $n^{\#(\gamma^{-1} \alpha)}$. In conclusion, the total contribution of loops is $k^{\# \alpha} n^{\#(\gamma^{-1} \alpha)}$. The contribution of the $X$ boxes is straightforward to compute, since these boxes are connected only by $\beta$. After the removal we get $\# \beta$ connected components of powers of $X$, and the total contribution is $\trace_\beta(X)$. 

Putting all this factors together, we obtain the following proposition:
\begin{proposition}
The mean $p$-th moment of the random matrix $Y = \trace_k[U X U^*]$ is given by
\begin{equation}\label{eq:moments_general}
\E[\trace(Y^p)] = \sum_{\alpha, \beta \in \S_p} k^{\# \alpha} n^{\#(\gamma^{-1} \alpha)} \trace_\beta(X) \Wg(\alpha \beta^{-1}).
\end{equation}
\end{proposition}

In the particular case where $X$ is a rank one projector, one has $\trace_\beta(X) = 1$ for all permutations $\beta$ and since 
	\[\sum_{\sigma \in \S_p} \Wg(nk, \sigma) = \left( \prod_{j=0}^{p-1}(nk+j)\right)^{-1},\]
one obtains the following simplification:
\begin{equation}
\E[\trace(Y^p)] = \left( \prod_{j=0}^{p-1}(nk+j)\right)^{-1}\sum_{\alpha \in \S_p}k^{\# \alpha} n^{\#(\gamma^{-1} \alpha)}.
\end{equation}
We will discuss at length the simplifications that occur when dealing with rank one projectors in the forthcoming paper \cite{CN3}.

\section{Tensor products of independent random quantum channels}\label{sec:product_independent}

In the present section and in the next one, we consider two different models of tensor products of random quantum channels. In both cases, we first fix an interesting input state $X_{12}$ (the Bell state) and investigate the random matrix
\[[\Phi_1 \otimes \Phi_2](X_{12}). \]
Our two models correspond to the choice of two different 
random pairs $(\Phi_1,\Phi_2)$.
In both models, the channels $\Phi_1,\Phi_2$ are defined by random unitary matrices $U_1,U_2$ via the Stinespring representation introduced in Eq. (\ref{eq:Stinespring_form}).

The difference between the two cases lies in the correlation between the random matrices  $U_1$ and $U_2$. In the first model, interaction unitaries $U_{1,2}$ are \emph{independent} Haar distributed random matrices. The second model, which is more involved, deals with the case where $U_1$ is distributed according to the Haar measure and $U_2 = \ol{U_1}$. This choice introduces an interesting symmetry into the problem and such a construction has become classical in quantum information theory \cite{hayden, hastings}. Asymptotic results in this case shed light on the interesting phenomenon that the output of the product channel has one ``large'' eigenvalue. We call this phenomenon the \emph{Bell state phenomenon}.
In order to simplify the notation, we shall assume that $Y_1$ and $Y_2$ are rank-one projectors and that $n_1 = n_2 = n$, $k_1 = k_2 = k$.

Before looking in detail at the two models of interest, let us make one brief comment on the choice of the input state of the channels. It is clear that if one chooses an input state which factorizes $X_{12} = X_1 \otimes X_2$, then 
\[[\Phi_1 \otimes \Phi_2](X_{12}) = \Phi_1(X_1) \otimes \Phi_2(X_2),\]
and there is no correlation (classical or quantum) between the channels. In order to avoid such trivial situations, one has to choose an input state which is \emph{entangled}. An obvious choice (given that $n_1 = n_2 = n$) is to take $X_{12} = E_n$, the $n$-dimensional Bell state (or the maximally entangled state), and we shall use this state in what follows.

The first model, although new, does not bring strikingly new results from the random matrix point of view. We treat it here as an illustration of what our calculus can allow to compute, and as a point of comparison with the second model.

\subsection*{Independent interaction unitaries}

In the remaining of this section, we consider two \emph{independent} realizations $U_1 = U$ and $U_2 = V$ of Haar-distributed unitary random matrices on $\U(nk)$. For both channels the state of the environment is a rank-one projector and we are interested in the $n^2 \times n^2$ random matrix
	\[Z = [\Phi^U \otimes \Phi^V] (E_n),\]
where $E_n$ is the maximal entangled Bell state (notice the $1/n$ normalization) 
\[E_n = \frac{1}{n} \sum_{i,j=1}^n \ketbra{e_i}{e_j} \otimes \ketbra{e_i}{e_j}.\]
	
\begin{figure}[ht]
\includegraphics{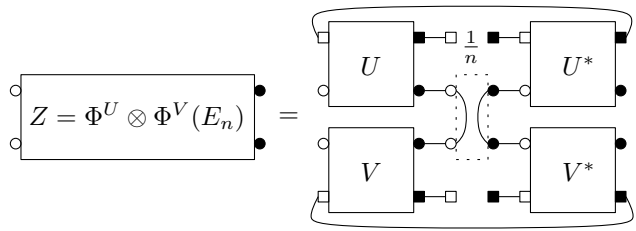}
\caption{$Z = \Phi^U \otimes \Phi^V (E_n)$}
\label{fig:bi_channel_UV}
\end{figure}

The diagram associated with the (2,2) tensor $Z$ is drawn in the Figure \ref{fig:bi_channel_UV}; we chose to represent by squares decorations corresponding to $\C^k$ and by circles decorations corresponding to $\C^n$.

As usual, we are interested in computing the moments $\E[\trace(Z^p)]$ for all $p \geq 1$ using the graphical method. We start by replacing $U^*$ (resp. $V^*$) blocks by $\ol U$ (resp. $\ol V$) blocks. An important point here is that there are two type of blocks corresponding to the independent random unitary matrices $U$ and $V$ (when computing the $p$-th moment of $Z$, there are $p$ blocks of each type). This has two important consequences: when expanding the diagram in order to compute the expectation of the trace, one can only pair $U$ blocks with $\ol U$ blocks and $V$ blocks with $\ol V$ blocks; ``cross-pairings'' between $U$ blocks and $V$ blocks are not allowed by the expansion algorithm. This algorithm proceeds iteratively, first by removing, say, the $U$ blocks (the $V$ blocks being treated as constants) and then by removing the $V$ blocks. Hence ``cross-pairings'' cannot occur. The second consequence of the presence of two independent Haar unitary matrices is that in the final expression for the expectation of the diagram, there will be two Weingarten weights, one for each independent unitary integration.

\begin{lemma}
The following holds true ($\gamma$ is the cycle permutation defined in Eq. (\ref{eq:gamma_cycle}))
\begin{equation}\label{eq:bi_canal_UV_det}
	\E[\trace(Z^p)] = \!\!\!\!\!\!\!\!\!\!\!\! \sum_{\alpha_U, \beta_U, \alpha_V, \beta_V \in \S_p} \!\!\!\!\!\!\!\!\!\!\!\! k^{\# \alpha_U + \# \alpha_V}n^{\#(\gamma^{-1}\alpha_U ) + \#(\gamma^{-1}\alpha_V ) + \# (\beta_U^{-1}\beta_V)-p}\Wg(\alpha_U\beta_U^{-1})\Wg(\alpha_V\beta_V^{-1}).
\end{equation}
\end{lemma}

\begin{proof}
As it has already been stated, the expectation with respect to both unitary matrices $U$ and $V$ can be seen as the result of two removal procedures, and hence the Weingarten sum shall be indexed by a pair of removals $(r_U,r_V)$. In other words, the Weingarten sum shall be indexed by 2 pairs of permutations, one for each type of block; we shall denote them by $\alpha_U, \beta_U, \alpha_V, \beta_V \in \S_p$. The four permutations are responsible for pairing blocks in the following way ($1 \leq i \leq p$):
\begin{enumerate}
	\item the white decorations of the $i$-th $U$-block are paired with the white decorations of the $\alpha_U(i)$-th $\ol U$ block;
	\item the black decorations of the $i$-th $U$-block are paired with the black decorations of the $\beta_U(i)$-th $\ol U$ block;
	\item the white decorations of the $i$-th $V$-block are paired with the white decorations of the $\alpha_V(i)$-th $\ol V$ block;
	\item the black decorations of the $i$-th $V$-block are paired with the black decorations of the $\beta_V(i)$-th $\ol V$ block.
\end{enumerate}
The diagram associated with $\trace(Z^p)$ contains, aside from the random unitary blocks, deterministic bras $\includegraphics{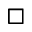}$ and kets $\includegraphics{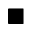}$. However, these boxes have a trivial contribution of 1 to the final result. Hence, the result of the graph expansion is a (sum over a) collection of loops, multiplied by some scalar factor. The different contributions of a quadruple $(\alpha_U, \beta_U, \alpha_V, \beta_V) \in \S_p^4 $ are given by (recall that circles correspond to $n$-dimensional spaces and squares correspond to $k$-dimensional spaces):
\begin{enumerate}
	\item ``$\includegraphics{square_w.eps}  U$''-loops: $k^{\# \alpha_U}$;
	\item ``$\includegraphics{square_w.eps}  V$''-loops: $k^{\# \alpha_V}$;
	\item ``$\includegraphics{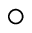}  U$''-loops: $n^{\#(\gamma^{-1}\alpha_U)}$;
	\item ``$\includegraphics{circle_w.eps}  V$''-loops: $n^{\#(\gamma^{-1}\alpha_V)}$;
	\item ``$\includegraphics{square_b.eps}  U$''-loops: none;
	\item ``$\includegraphics{square_b.eps}  V$''-loops: none;
	\item ``$\includegraphics{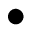}  U,V$''-loops: $n^{\# (\beta_U^{-1}\beta_V)}$;	
	\item normalization factors $1/n$ from the Bell matrices $E_n$: $n^{-p}$;
	\item Weingarten weights for the $U$-matrices: $\Wg(\alpha_U\beta_U^{-1})$;
	\item Weingarten weights for the $V$-matrices: $\Wg(\alpha_V\beta_V^{-1})$.
\end{enumerate}
Adding all these contributions, we obtain the announced exact closed-form expression.
\end{proof}

\subsection*{Asymptotics}

The preceding expression is intractable at fixed $n$ and $k$, so we study two asymptotic regimes: 
\begin{enumerate}
	\item[(I)] $n$ fixed, $k \to \iy$;
	\item[(II)] $k$ fixed, $n \to \iy$;
\end{enumerate}

At this stage, before looking into each particular asymptotic regime, we can make an important observation. Notice that in the preceding expression, aside from the factor $n^{\# (\beta_U^{-1}\beta_V)}$, the general term of the sum factorizes into a ``$(\alpha_U, \beta_U)$'' part and a ``$(\alpha_V, \beta_V)$'' part. This was to be expected, since the coupling between the two channels is realized by the input state $E_n$ which has a contribution of $n^{\# (\beta_U^{-1}\beta_V)-p}$. Let us also note that a third interesting asymptotic regime, $n,k \to \iy$, $k/n \to c >0$ will be studied using the same methods in a forthcoming paper.

\begin{theorem}
In the firs regime, $n$ fixed, $k \to \iy$, the output of the tensor product of the channels, for large values of $k$, is close to the chaotic state \[\rho_* = \frac{\I_{n^2}}{n^2}.\]
In the second regime, $k$ fixed, $n \to \iy$, the asymptotic eigenvalues of $Z$ are $1/k^2$ with multiplicity $k^2$ and $0$ with multiplicity $n^2 - k^2$.
\end{theorem}

\begin{proof}
Using the standard asymptotics for the Weingarten functions 
\begin{align*}
\Wg(\alpha_U\beta_U^{-1}) &\stackrel{k \to \iy}{\sim} (nk)^{-p-|\alpha_U\beta_U^{-1}|} \Mob(\alpha_U\beta_U^{-1}) \text{ and}\\
\Wg(\alpha_V\beta_V^{-1}) &\stackrel{k \to \iy}{\sim} (nk)^{-p-|\alpha_V\beta_V^{-1}|} \Mob(\alpha_V\beta_V^{-1}),
\end{align*}
the power of $k$ appearing in a general $(\alpha_U, \beta_U, \alpha_V, \beta_V)$ term is 
\[k^{p-|\alpha_U| + p-|\alpha_V| - p-|\alpha_U\beta_U^{-1}| - p-|\alpha_V\beta_V^{-1}|} = k^{-(|\alpha_U| + |\alpha_V| + |\alpha_U\beta_U^{-1}| + |\alpha_V\beta_V^{-1}|)}.\]
It is obvious that all the terms converge to zero, except the one with $\alpha_U = \beta_U = \alpha_V = \beta_V = \id$. Using $\Mob(\id) = 1$, we conclude that
	\[\lim_{k \to \iy} \E[\trace(Z^p)] = n^{2-2p}.\]
One can restate this in terms of the empirical eigenvalue distribution of the $n^2 \times n^2$ matrix $Z$:
	\[\mu_Z = \frac{1}{n^2}\sum_{i=1}^{n^2}\delta_{\lambda_i(Z)} \invstackrel{\longrightarrow}{k \to \iy} \delta_{1/n^2}.\]
In other words, the output of the tensor product of the channels, for large values of $k$, is close to the chaotic state
 \[\rho_* = \frac{\I_{n^2}}{n^2}.\]

As for the second regime, using similar considerations, we obtain
	\[\lim_{n \to \iy} \E[\trace(Z^p)] = k^{2-2p}.\]

\end{proof}

Note that both regimes presented here are trivial to some extent. We could prove at only a small additional cost that the convergence of the  eigenvalues is actually almost sure. See the next section for the technique of proof, a direct adaptation of the Borel-Cantelli lemma. 

Finally, we want to emphasize that the asymptotic behavior of the output in the second regime changes drastically when considering the conjugate case, and that this fact will have very important consequences in the theory of quantum information.

\section{Tensor products of conjugate random quantum channels}\label{sec:product_conjugate}

We have seen that tensor products of independent random channels 
have an eigenvalue behavior close to the single channel case (see \cite{nechita} for the treatment of the single channel case) - despite the fact that
the input state is maximally entangled. 
In this section, we consider the case where $U_1 = U$, $U_2 = \ol U$ and $U \in \U(nk)$ is a Haar uniform random unitary matrix. Tensor products of channels of this type are now classical in the literature (see \cite{hayden, hastings}). One of the reasons why channels of this particular form receive such attention is that one can show that the product channel has a ``trivial large eigenvalue'' of order $1/k$. We shall provide a graphical proof of this fact later, in Lemma \ref{lem:hayden}.

\begin{figure}[ht]
\includegraphics{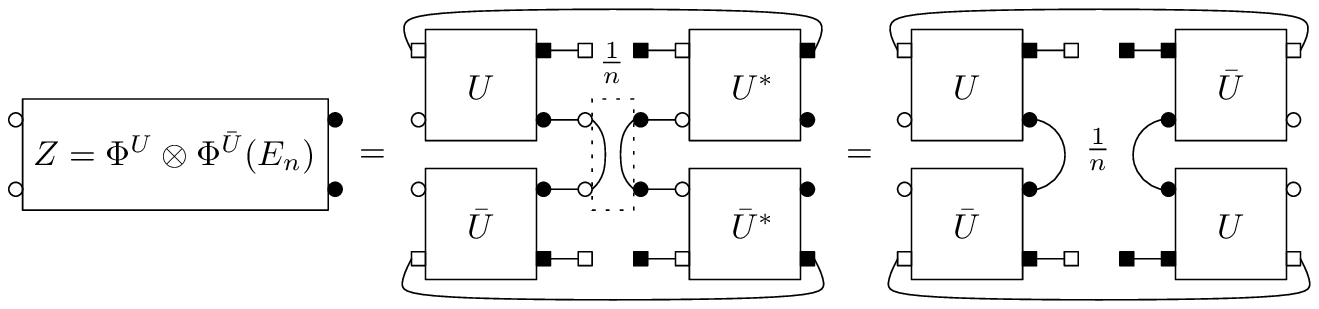}
\caption{$Z = \Phi^U \otimes \Phi^{\ol U} (E_n)$}
\label{fig:bi_channel_UUbar}
\end{figure}

Again, we are interested in the moments of the random matrix $Z = \Phi^U \otimes \Phi^{\ol U} (E_n)$, depicted in the Figure \ref{fig:bi_channel_UUbar}.

This time calculations are more complicated, because only one unitary matrix appears in the product channel. This means that in the removal algorithm, one can pair boxes from $\Phi^U$ with boxes from $\Phi^{\ol U}$ thus obtaining more complicated patterns. Another consequence of the fact that we use only one random unitary matrix is that the Weingarten sums are indexed by only one pair of permutations $(\alpha, \beta) \in \S_{2p}^2$.

In order to count the loops obtained after the graph expansion, we label the $U$ and the $\ol U$ boxes in the following manner: $1^T, 2^T, \ldots, p^T$ for the $U$ boxes of the first channel (T as ``top'') and $1^B, 2^B, \ldots, p^B$ for the $U$ boxes of the second channel (B as ``bottom''). We shall also order the labels as $\{1^T, 2^T, \ldots, p^T, 1^B, 2^B, \ldots, p^B\} \isom \{1, \ldots, 2p\}$. A removal $r=(\alpha, \beta) \in \S_{2p}^2$ of the random ($U$ and $\ol U$)  boxes connects the decorations in the following way:
\begin{enumerate}
	\item the white decorations of the $i$-th $U$-block are paired with the white decorations of the $\alpha(i)$-th $\ol U$ block;
	\item the black decorations of the $i$-th $U$-block are paired with the black decorations of the $\beta(i)$-th $\ol U$ block.
\end{enumerate}
Next, we introduce two fixed permutations $\gamma, \delta \in \S_{2p}$ which will be useful in counting the loops. The permutation $\gamma$ represents the initial wiring of the $\includegraphics{circle_w.eps}$ decorations (before the graph expansion) and $\delta$ accounts for the wires between the $\includegraphics{circle_b.eps}$ decorations (which come from $E_n$).
More precisely, for all $i$,
\begin{equation}
\label{eq:def-gamma-delta}
\gamma(i^T) = (i-1)^T, \quad \gamma(i^B) = (i+1)^B,\quad
and \quad
\delta(i^T) = i^B, \quad \delta(i^B) = i^T.
\end{equation}

We are now ready to compute the average moments of the random matrix $Z$.
\begin{lemma}
The following holds true
\begin{equation}\label{eq:bi_canal_UUbar_det}
	\E[\trace(Z^p)] = \sum_{\alpha, \beta \in \S_{2p}}k^{\# \alpha}n^{\#(\alpha \gamma^{-1}) +\# (\beta\delta) - p}\Wg(\alpha\beta^{-1}).
\end{equation}
\end{lemma}

\begin{proof}
With the notations introduced above, we can now count the contributions for each individual pairing $(\alpha, \beta)$:
\begin{enumerate}
	\item ``$\includegraphics{square_w.eps}$''-loops: $k^{\# \alpha}$;
	\item ``$\includegraphics{circle_w.eps}$''-loops: $n^{\#(\alpha \gamma^{-1})}$;	
	\item ``$\includegraphics{square_b.eps}$''-loops: none;
	\item ``$\includegraphics{circle_b.eps}$''-loops: $n^{\# (\beta\delta^{-1})} = n^{\# (\beta\delta)}$ (notice that $\delta$ is an involution);
	\item normalization factors $1/n$ from the $E_n$ matrices: $n^{-p}$;
	\item Weingarten weights for the $U$-matrices: $\Wg(\alpha\beta^{-1})$.
\end{enumerate}

Adding up all the above contributions, we obtain the claimed formula.
\end{proof}

In the rest of the section, we shall focus on the asymptotic regime $n$ fixed, $k \to \iy$, and in the next section we shall look into the more interesting case  $k$ fixed, $n \to \iy$.

Before stating the asymptotic result, let us make two preliminary remarks. In the case of the conjugate product channel, since only one unitary matrix appears in the diagrams, there is a notable difference concerning the asymptotics of the Weingarten function:
\[\Wg(\alpha\beta^{-1}) \sim (nk)^{-2p - |\alpha\beta^{-1}|}\Mob(\alpha\beta^{-1}).\]
One can easily compute the following quantities involving the permutations $\gamma$ and $\delta$ which will be useful later, when doing asymptotics: $|\gamma| = 2p-2$, $|\delta| = p$, $|\gamma \delta| = p$.

\begin{proposition}
In the asymptotic  regime where $n$ is fixed and $k \to \iy$, the random matrix $Z$ converges to the chaotic state 
 \[\rho_* = \frac{\I_{n^2}}{n^2}.\]
\end{proposition}

\begin{proof}
Computing the asymptotic trace for large $k$ gives
\[\E[\trace(Z^p)] \sim \sum_{\alpha, \beta \in \S_{2p}} k^{-(|\alpha|+|\alpha\beta^{-1}|)}n^{p-(|\alpha\gamma^{-1}|+|\beta\delta|+|\alpha\beta^{-1}|)}\Mob(\alpha\beta^{-1}).\]

Minimizing the power of $k$ above gives $|\alpha|+|\alpha\beta^{-1}| \geq 0$, with equality iff $\alpha = \beta = \id$, hence
\[\lim_{k \to \iy} \E[\trace(Z^p)] = n^{2-2p},\]
and the conclusion is the same as in the case of two independent quantum channels: the output $Z$ is asymptotically close to the chaotic state $\rho_*$.
\end{proof}

\subsection{Conjugate channels, the Bell phenomenon}

We are left with studying the most interesting regime, $k$ fixed and $n \to \iy$. Our main result is as follows:

\begin{theorem}
\label{thm:bell-phenomenon}
In the regime of $k$ fixed, $n \to \iy$, the eigenvalues of the matrix $Z$ converge \emph{almost surely} towards:
\begin{itemize}
\item $\frac{1}{k} + \frac{1}{k^2} - \frac{1}{k^3}$, with multiplicity one;
\item $\frac{1}{k^2} - \frac{1}{k^3}$, with multiplicity $k^2-1$;
\item $0$, with multiplicity $n^2 - k^2$.
\end{itemize}
\end{theorem}

Note that it follows from the Stinespring theorem that $Z$ has at most $k^2$ non-zero eigenvalues. Therefore a moment approach is possible for the proof.

We start with a technical Lemma about the structure of geodesics between 
the specific permutations
$\gamma$ and $\delta$ introduced in Equation (\ref{eq:def-gamma-delta}).

\begin{lemma}
\label{lem:elementary}
For $1 \leq i \leq p$, let $\tau_i$ be the transposition $\left(i^T \;,\; (i-1)^B\right)$. Then the permutations $\alpha$ on the geodesic $\gamma \to \alpha \to \delta$ are indexed by subsets $A \subseteq \{1, \ldots, p\}$: $\alpha = \gamma \prod_{i \in A} \tau_i$. Moreover, for such a permutation, we have
\[
|\alpha| = 
\begin{cases}
2p-2 & \text{if $A=\emptyset$,}\\
2p-|A|& \text{if $A \neq \emptyset$.}
\end{cases}
\]
\end{lemma}
\begin{proof}
If $A = \emptyset$, then $|\alpha| = |\gamma| = 2p-2$. Otherwise, after computing the action of $\alpha$
\begin{align*}
\alpha(i^T) &= 
\begin{cases}
i^B & \text{ if $i \in A$,}\\
(i-1)^T & \text{ if $i \notin A$;}\\
\end{cases}\\
\alpha(i^B) &= 
\begin{cases}
i^T & \text{ if $(i+1) \in A$,}\\
(i+1)^B & \text{ if $(i+1) \notin A$;}\\
\end{cases}
\end{align*}
it is easy to see that each element $i$ of $A$ spans a cycle of $\alpha$ and thus $|\alpha| = 2p - \#\alpha = 2p-|A|$.
\end{proof}

We split this proof of Theorem \ref{thm:bell-phenomenon} in two steps: first we prove the convergence in expectation, and then
we prove the almost sure convergence. 

\begin{proof}[Proof of the convergence in expectation]
Using the same asymptotic formula as in the previous section (this time for large $n$), the quantity one wants to minimize in this case is
	\[|\alpha \gamma^{-1}| + |\beta \delta| + |\alpha \beta^{-1}| = |\gamma^{-1}\alpha | + |\alpha^{-1} \beta| + |\beta^{-1} \delta| \geq |\gamma^{-1} \delta| = p.\]
Equality is attained when $\gamma \to \alpha \to \beta \to \delta$ is a geodesic in $\S_{2p}$. Using this observation, we obtain
\[\E[\trace(Z^p)] \sim \sum_{\gamma \to \alpha \to \beta \to \delta} k^{-(|\alpha|+|\alpha\beta^{-1}|)}\Mob(\alpha\beta^{-1}).\]

It turns out that we can compute exactly the last sum as follows. First, notice that the geodesic condition $\gamma \to \alpha \to \beta \to \delta$ can be restated as $\id \to \gamma^{-1}\alpha \to \gamma^{-1}\beta \to \gamma^{-1}\delta$. But $\gamma^{-1}\delta$ is a product of $p$ transpositions with disjoint support: $\gamma^{-1}\delta = \prod_{i=1}^p \tau_i$, where $\tau_i$ is the transposition $\left(i^T \;,\; (i-1)^B\right)$ for all $1 \leq i \leq p$. As in Lemma \ref{lem:elementary}, permutations on a geodesic between $\id$ and $\gamma^{-1}\delta$ are parameterized by subsets of $\{1, \ldots, p\}$ as follows. Permutations $\gamma^{-1}\alpha$ and $\gamma^{-1}\beta$ lie on a geodesic between $\id$ and $\gamma^{-1}\delta$ (i.e. $\id \to \gamma^{-1}\alpha \to \gamma^{-1}\beta \to \gamma^{-1}\delta$) if and only if there exist two subsets $\emptyset \subseteq A \subseteq B  \subseteq \{1, 2, \ldots, p\}$ such that
\begin{align*}
\gamma^{-1}\alpha &= \prod_{i \in A} \tau_i,\\
\gamma^{-1}\beta &= \prod_{i \in B} \tau_i.
\end{align*}
For two such permutations, it is obvious that $|\alpha^{-1} \beta| = |(\gamma^{-1}\alpha)^{-1} \gamma^{-1}\beta| = |B \setminus A|$. 
In order to compute  $|\alpha|$, we rely on Lemma \ref{lem:elementary}.

Since $\alpha^{-1}\beta$ is a product of $|B \setminus A|$ transpositions of disjoint support, it follows by \cite{collins-sniady} that $\Mob(\alpha^{-1}\beta) = (-1)^{|B \setminus A|}$ and we are left with the following expression:
\[\E[\trace(Z^p)] \sim \sum_B k^{-(2p-2 +|B|)} (-1)^{|B|} + \sum_{\emptyset \neq A \subseteq B} k^{-(2p-|A| + |B \setminus A|)} (-1)^{|B \setminus A|}.\]
Using the multinomial identities
\begin{align*}
\sum_{\emptyset \subseteq A \subseteq \{1, \ldots, p\}} \!\!\!\! x^{|A|} &= (1+x)^p \quad \text{and}\\
\sum_{\emptyset \subseteq A \subseteq B \subseteq \{1, \ldots, p\}} \!\!\!\!\!\!\!\! x^{|A|}y^{|B \setminus A|} &= (1+x+y)^p,
\end{align*}
we obtain the asymptotic traces of the output matrix $Z$:
\[\E[\trace(Z^p)]  \sim  \left( \frac{1}{k} + \frac{1}{k^2} - \frac{1}{k^3} \right)^p + (k^2-1) \left(\frac{1}{k^2}-\frac{1}{k^3} \right)^p.\]

We conclude that the matrix $Z$ has, asymptotically, the following eigenvalues:
\begin{itemize}
\item $\frac{1}{k} + \frac{1}{k^2} - \frac{1}{k^3}$, with multiplicity one;
\item $\frac{1}{k^2} - \frac{1}{k^3}$, with multiplicity $k^2-1$;
\item $0$, with multiplicity $n^2 - k^2$.
\end{itemize}

\end{proof}

Next we move on to the proof of almost sure convergence. 
We would like to mention about the proof below
that we believe that it should be possible 
to prove that
$$ \E\left[ \left(\trace(Z^p) - \E\trace(Z^p)\right)^2 \right] =O(n^{-2})$$
simply by observing that the function 
$$U\to \trace(Z^p)$$
is Lipschitz on the unitary group and by applying a 
Gromov-Milman type concentration measure argument. We refer
 to \cite{hayden-leung-winter} for an exposition of such techniques. 
 The authors acknowledge that this approach might be slightly 
 less cumbersome in the specific case of this proof. 
 However, we chose to keep our proof of a combinatorial nature
 for the sake of coherence.

\begin{proof}[Proof of the almost sure convergence]
It is a standard technique in probability theory that in order to show the almost-sure convergence of the eigenvalues of $Z$ to their respective limits, it suffices for the covariance series to converge, for all values of $p$:
\[\sum_{n=1}^\iy \E\left[ \left(\trace(Z^p) - \E\trace(Z^p)\right)^2 \right] < \iy.\]
Indeed, this inequality together with the Borel-Cantelli lemma imply that
almost surely as $n\to\infty$,
$$\trace(Z^p) \to \E\trace(Z^p)$$

The two 
ingredients which make the proof work are the following. The first one is the fact that the error one makes when approximating the Weingarten function with its dominating term is of the order $-2$:
\[\Wg(\alpha) = (nk)^{-(p + |\alpha|)} (\Mob(\alpha) + O((nk)^{-2})).\]
This follows from Theorem \ref{thm:mob} and the definition of $\Mob$ below.
The second ingredient is contained in the geodesic inequality $|\gamma^{-1}\alpha | + |\alpha^{-1} \beta| + |\beta^{-1} \delta| \geq |\gamma^{-1} \delta| = p$. Earlier, we have completely described the set of couples $(\alpha, \beta)$ which saturate the equality. It turns out that one can say more on the values of the function $(\alpha, \beta) \mapsto E(\alpha, \beta) = |\gamma^{-1}\alpha | + |\alpha^{-1} \beta| + |\beta^{-1} \delta| -p$, as follows. 
Applying Lemma \ref{lem:S_p} two times, it is clear that the values taken by $E(\alpha, \beta)$ are all even, $E(\alpha, \beta) \neq 1$ and thus 
\begin{align*}
\E[\trace(Z^p)] &= \sum_{\alpha, \beta \in \S_{2p}}k^{\# \alpha}n^{\#(\alpha \gamma^{-1}) +\# (\beta\delta) - p}\Wg(\alpha\beta^{-1})\\
&=\sum_{\gamma \to \alpha \to \beta \to \delta} k^{-(|\alpha| + |\alpha \beta^{-1}|)}\Mob(\alpha \beta^{-1}) + O\left( \frac{1}{n^2}\right).
\end{align*}

We start by computing the easiest term of the covariance, namely
\[\E[\trace(Z^p)]^2 = \!\!\!\!\!\!\!\! \sum_{\substack{\gamma \to \alpha_1 \to \beta_1 \to \delta \\ \gamma \to \alpha_2 \to \beta_2 \to \delta}} \!\!\!\!\!\!\!\! k^{-(|\alpha_1| + |\alpha_1 \beta_1^{-1}| + |\alpha_2| + |\alpha_2 \beta_2^{-1}|)}\Mob(\alpha_1 \beta_1^{-1})\Mob(\alpha_2 \beta_2^{-1}) + O\left( \frac{1}{n^2}\right).\]

The second term $\E[\trace(Z^p)^2]$ is more difficult to estimate and one needs to introduce the  permutations $\bar \gamma,\bar \delta \in \S_{4p}$ :
\begin{align*}
\bar \gamma &= (1^T 2^T \cdots p^T)((p+1)^T (p+2)^T \cdots (2p)^T)(1^B 2^B \cdots p^B)((p+1)^B (p+2)^B \cdots (2p)^B) ;\\
\bar \delta &= (1^T 1^B) (2^T 2^B) \cdots (p^T p^B) ((p+1)^T (p+1)^B) \cdots ((2p)^T (2p)^B).
\end{align*}
With this notation, we have 
\[\E[\trace(Z^p)^2] = \sum_{\alpha, \beta \in \S_{4p}}k^{\# \alpha}n^{\#(\alpha \bar \gamma^{-1}) + \#(\beta \bar \delta) - p}\Wg(\alpha \beta^{-1}) =\]
\[\sum_{\alpha, \beta \in \S_{4p}} k^{-(|\alpha| + |\alpha \beta^{-1}|)}n^{2p-(|\bar \gamma^{-1}\alpha |+|\alpha^{-1} \beta |+|\beta^{-1} \bar \delta|)}\Mob(\alpha \beta^{-1}) + O\left( \frac{1}{n^2}\right).\]

One can easily show that $|\bar \gamma^{-1}\alpha |+|\alpha^{-1} \beta |+|\beta^{-1} \bar \delta| \geq |\bar \gamma^{-1}\bar \delta| = 2p$. Since both $\bar \gamma$ and $\bar \delta$ leave invariant the sets $\{1^{T,B}, 2^{T,B}, \ldots, p^{T,B}\}$ and $\{(p+1)^{T,B}, (p+2)^{T,B}, \ldots, 2p^{T,B}\}$, geodesic couples $(\alpha, \beta)$ are obtained as direct sums
\begin{align*}
\alpha &= \alpha_1 \oplus \alpha_2,\\
\beta &= \beta_1 \oplus \beta_2,
\end{align*}
where $\alpha_1, \beta_1 \in \S(\{1^{T,B}, 2^{T,B}, \ldots, p^{T,B}\})$, $\alpha_2 \beta_2 \in \S(\{(p+1)^{T,B}, (p+2)^{T,B}, \ldots, 2p^{T,B}\})$ are such that $\gamma_1 \to \alpha_1 \to \beta_1 \to \delta_1$ and $\gamma_2 \to \alpha_2 \to \beta_2 \to \delta_2$ are geodesics (the permutations $\gamma_{1,2}$ and $\delta_{1,2}$ are defined in an obvious way). One has also that $|\alpha|= |\alpha_1| + |\alpha_2|$, $|\beta|= |\beta_1| + |\beta_2|$ and that $\Mob(\alpha \beta^{-1}) = \Mob(\alpha_1 \beta_1^{-1}) \Mob(\alpha_2 \beta_2^{-1})$. Putting all this together, one gets the final expression which ends the proof
\[\E[\trace(Z^p)^2] = \sum_{\substack{\gamma \to \alpha_1 \to \beta_1 \to \delta \\ \gamma \to \alpha_2 \to \beta_2 \to \delta}}k^{-(|\alpha_1| + |\alpha_1 \beta_1^{-1}| + |\alpha_2| + |\alpha_2 \beta_2^{-1}|)}\Mob(\alpha_1 \beta_1^{-1})\Mob(\alpha_2 \beta_2^{-1}) + O\left( \frac{1}{n^2}\right).\]

\end{proof}

\subsection{Generalization of Theorem \ref{thm:bell-phenomenon} and
discussion}

\begin{figure}[ht]
\includegraphics{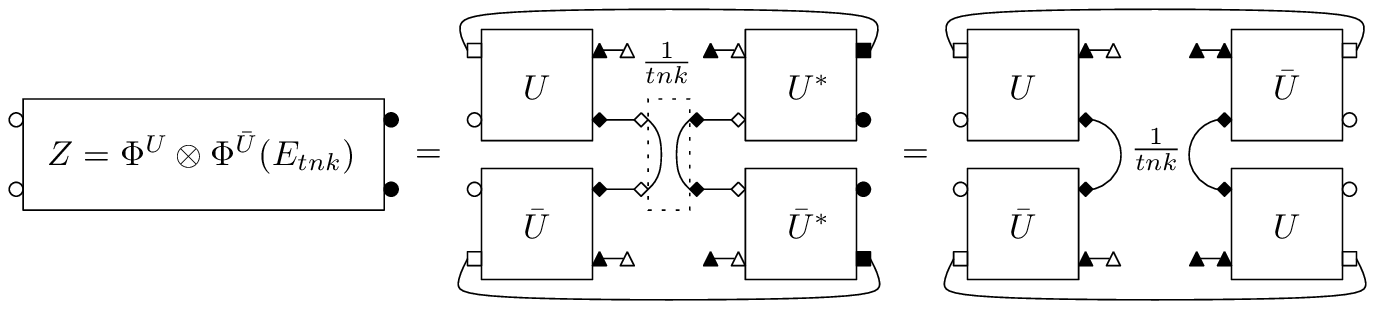}
\caption{$Z = \Phi^U \otimes \Phi^{\ol U} (E_{tnk})$}
\label{fig:bi_channel_UUbarmodifie}
\end{figure}

We finish this paper by studying a generalization of the model investigated in the previous section. We consider
the case where $k$ is a fixed integer, and
$t\in (0,1)$ is a fixed number (possibly a function of $k$). 
For each $n$, we consider a random unitary matrix $U \in \M_{nk}(\C )$, and a projection $q_n$ of $\M_{nk} (\C )$ of rank
$p_n$ such that $p_n/(nk)\sim t$ as $n\to\infty$.
Our model of a random quantum channel is 
$$\Phi : \M_{p_n}(\C )\to \M_n(\C )$$
given by
$$\Phi (X)=\trace_k (U X U^*)$$
where the density matrix $X$ satisfies $X\leq q_n$ (in other words we consider the isomorphism $q_n\M_{nk}(\C )q_n \simeq \M_{p_n}(\C )$). Graphically, our model amounts to Figure \ref{fig:bi_channel_UUbarmodifie}.

As usual, we are interested in the process given by the eigenvalues of $Z$ as $n\to\infty$ (in our setup, $k$ is fixed). Here, with almost the same techniques as in Theorem \ref{thm:bell-phenomenon}, we obtain the following result.

\begin{theorem}
\label{thm:product-channel}
Almost surely, as $n \to \iy$, the random matrix $\Phi\otimes\overline\Phi(E_{tnk}) \in \M_{n^2}(\C)$ has non-zero eigenvalues converging towards
\[\gamma^{(t)} = \left( t + \frac{1-t}{k^2},\underbrace{\frac{1-t}{k^2}, \ldots, \frac{1-t}{k^2}}_{k^2-1 \text{ times}}\right).\]
\end{theorem}

\begin{proof}
Theorem 
\ref{thm:bell-phenomenon}
 is a particular 
case of this theorem with $t =(1-1/k)$
and it has been worked out in great detail.
Therefore we leave the reader to work out the appropriate modifications
 to this case.
\end{proof}

The striking fact here is that the largest eigenvalue behaves almost surely
like $ t + (1-t)/k^2$. 
The existence of a large eigenvalue was already anticipated 
by Hayden in \cite{hayden}, Lemma II.2. 
The lemma below is a slight generalization of Hayden's lemma, following his idea.

\begin{lemma}
\label{lem:hayden}
In the above model, the largest eigenvalue is at least $t$.
\end{lemma}

For the sake of being self contained, we give a proof of this fact. Moreover, the proof is graphical, using our diagrammatic calculus developed in Section \ref{sec:graphic-model}.

\begin{proof}
Following \cite{hayden}, it is enough to prove that
\[\trace(Z_nE_n) = \scalar{\frac{1}{\sqrt n}\Bell_n}{\left[\Phi^U\otimes \Phi^{\overline{U}}(E_{tkn})\right] \frac{1}{\sqrt n}\Bell_n}\geq t.\]

In order to accomplish this, we use the diagram invariance to stretch outside the inner parts of the diagram and then we notice that the resulting diagram is of the form $\scalar{X}{X}$ for some $X \in \C^{k^2}$, see Figure \ref{fig:largest_eig_A}.
\begin{figure}[ht]
\includegraphics{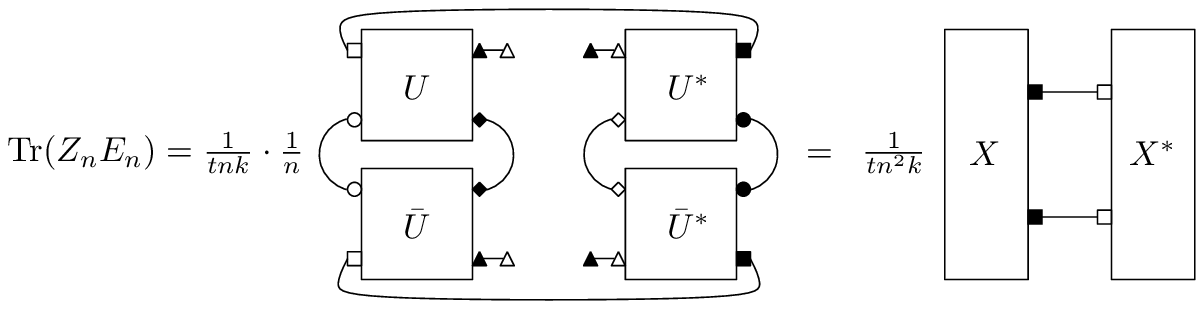}
\caption{Re-writing $\trace(Z_nE_n)$}
\label{fig:largest_eig_A}
\end{figure}
Introducing the orthogonal projector $A = \I_{k^2} - E_k \in \M_{k^2}(\C)$, it is obvious that $\scalar{X}{AX} \geq 0$ and thus 
the inequality in Figure \ref{fig:largest_eig_C} holds. Note that we have replaced the identity operator $\I_{k^2}$ connecting $X$ and $X^*$ by the maximally entangled state on $\C^k \otimes \C^k$. 
\begin{figure}[ht]
\includegraphics{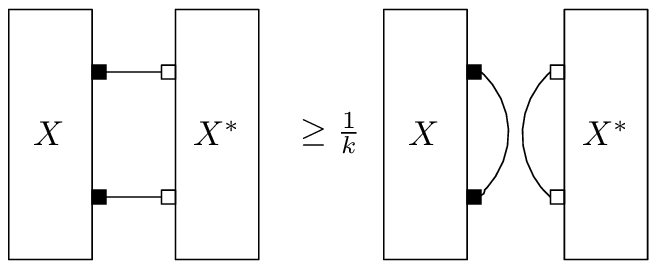}
\caption{Replacing $\I_{k^2}$ by $E_k$}
\label{fig:largest_eig_C}
\end{figure}

Now we use the unitary axioms on each of the two connected diagrams above and we obtain the result in Figure \ref{fig:largest_eig_D}.
\begin{figure}[ht]
\includegraphics{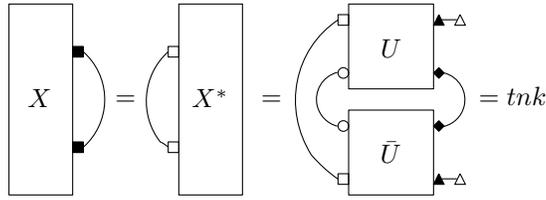}
\caption{Application of the unitary axioms}
\label{fig:largest_eig_D}
\end{figure}

Putting all the factors together, we get
\[\trace(Z_nE_n) \geq \frac{1}{tn^2k^2}(tnk)^2 = t,\]
and then, since $E_n$ is an orthogonal projector, we conclude that the largest eigenvalue of $Z_n$ is at least $t$.
\end{proof}

To conclude, let us compare Theorem 
\ref{thm:product-channel} and Lemma \ref{lem:hayden}.
Independently on the choice of $t$ and $k$, 
the value that we obtain almost surely for the largest eigenvalue in Theorem \ref{thm:product-channel}
 improves strictly the lower bound for the largest eigenvalue obtained in Lemma
 \ref{lem:hayden}, as $t+(1-t)/k^2>t$.
 
 For fixed $t$, the relative improvement $(t+(1-t)/k^2)/t$ becomes small as $k$ becomes big. 
 On the other hand, if $t\leq k^{-2}$, the Lemma \ref{lem:hayden}
 does not bring new information, as the largest eigenvalue is always
 at least $k^{-2}$, whereas Theorem \ref{thm:product-channel}
 brings new information. 
 
  So, the relative improvement is optimal for $t\sim k^{-2}$.
 This heuristic study leads us to think that it is possible to improve 
known counterexamples about the various entropy
additivity conjectures, and this is the object of study of our second paper \cite{CN2}.

\section*{Acknowledgments}

One author
(B.C.) would like to thank P. Hayden for an enlightening talk and
conversations in Guadalajara during the summer 2007. 
This is where he discovered random quantum channels and
additivity problems. 
He is also grateful to the audiences of two preliminary lectures
about these results in Sendai and Yokohama.
He also thanks T. Hayashi, R. Burstein and P. \'Sniady
for discussions at various stages about related topics.

This research was conducted partly in Lyon 1 and Ottawa. The authors
are grateful to these institutions for hosting their research.
One author (I. N.) benefited from funding by the conference 
``Random matrices, related topics and applications'' 
at C.R.M. and the mini-workshop
``Introduction to infinite-dimensional topological groups", 
 in Ottawa organized by M. Neufang. He thanks the organizers
of these two events for making a first visit to eastern Canada possible.
B.C.'s research was partly funded by ANR GranMa and ANR Galoisint.
The research of both authors was supported in part by NSERC
grant RGPIN/341303-2007.


\begin{thebibliography}{99}


\bibitem{coecke} 
Coecke, B. 
{\it Kindergarten quantum mechanics --- lecture notes}   Quantum theory: reconsideration of foundations---3,  81--98, AIP Conf. Proc., 810, Amer. Inst. Phys., Melville, NY, 2006. 

\bibitem{collins-imrn}
Collins, B.
{\it Moments and Cumulants of Polynomial random variables on unitary groups, 
the Itzykson-Zuber integral and free probability }
Int. Math. Res. Not., (17):953-982, 2003. 

\bibitem{CN2}
Collins, B. and Nechita, I. {\it Random quantum channels II: Entanglement of random subspaces, Renyi entropy estimates and additivity problems.} arXiv:0906.1877

\bibitem{CN3}
Collins, B. and Nechita, I. {\it Random Quantum Channels III}

\bibitem{collins-sniady}
Collins, B. and \'Sniady, P. {\it Integration with respect to the Haar measure on unitary, orthogonal and symplectic group.} Comm. Math. Phys. 264 (2006), no. 3, 773--795. 

\bibitem{hastings}
Hastings, M.B.
{\it Superadditivity of communication capacity using entangled inputs}. Nature Physics 5, 255 - 257 (2009).

\bibitem{hayden}
Hayden, P. {\it The maximal p-norm multiplicativity conjecture is false}. arXiv/0707.3291v1.

\bibitem{hayden-leung-winter}
Hayden, P., Leung, D.  and Winter, A. 
{\it Aspects of generic entanglement}. Comm. Math. Phys. 265 (2006), 95--117. 

\bibitem{hayden-winter}
Hayden, P. and Winter, A. 
{\it Counterexamples to the maximal p-norm multiplicativity conjecture for all $p>1$}. 
Comm. Math. Phys. 284 (2008), no. 1, 263--280.

\bibitem{jones}
Jones, V.F.R.
{\it Planar Algebras}
 arXiv:math/9909027v1

\bibitem{nechita}
Nechita, I. {\it Asymptotics of random density matrices.} Ann. Henri Poincar\'e 8 (2007), no. 8, 1521--1538. 

\bibitem{nica-speicher}
Nica, A and Speicher, R. {\it Lectures on the combinatorics of free probability}, volume 335 of London Mathematical Society Lecture Note Series. Cambridge University Press, Cambridge, 2006.

\end{thebibliography}
\end{document}